\documentstyle[11pt,amssymb,amsfonts,amsthm,amssymb,amscd,amstext,epsfig]{article}   
  
\def\ben{\begin{equation}}
\def\een{\end{equation}}

\let\a=\alpha \let\b=\beta  \let\d=\delta \let\e=\varepsilon
  \let\q=\theta 
\let\l=\lambda     
 \let\t=\tau  

\let\w=\omega \let\G=\Gamma

\let\pa=\partial
\def\be{\begin{equation}}
\def\ee{\end{equation}}
\def\ba{\begin{array}}
\def\ea{\end{array}}

\def\dalemb#1#2{{\vbox{\hrule height .#2pt
        \hbox{\vrule width.#2pt height#1pt \kern#1pt
                \vrule width.#2pt}
        \hrule height.#2pt}}}

\newcommand{\bea}{\begin{eqnarray}}
\newcommand{\eea}{\end{eqnarray}}

\newcommand{\Tr}{{\rm Tr} }

\def\R{{{\Bbb R}}}

\def\Im{{{\frak{Im}}}}

\def\Lag{{\mathcal{L}}}
\def\En{{\mathcal{G}}}

\begin{document}
\begin{flushright}
\hfill{DAMTP-2005-73} \\
{hep-th/0508092}
\end{flushright}

\begin{center}
\vspace{1cm} { \LARGE {\bf AdS black holes and thermal Yang-Mills 
correlators}}

\vspace{1.5cm}

Sean A. Hartnoll${}^1$ and S. Prem Kumar${}^{1,2}$

\vspace{0.8cm}

{\it ${}^1$ DAMTP, Centre for Mathematical Sciences,
Cambridge University\\ Wilberforce Road, Cambridge CB3 OWA, UK}

\vspace{0.3cm}

{\it ${}^2$ Department of Physics, University of Wales Swansea\\
Swansea, SA2 8PP, UK}

\vspace{0.3cm}

s.a.hartnoll@damtp.cam.ac.uk \hspace{1cm} p.kumar@damtp.cam.ac.uk

\vspace{2cm}

\end{center}

\begin{abstract}

We study the real time correlators of scalar glueball operators for
Yang-Mills theory at finite temperature in flat space.
The analytic structure of the 
frequency space propagator in perturbative field
theory is seen to be qualitatively different to the strong coupling
results that may be obtained from perturbations about AdS black hole
spacetimes: we find branch cuts rather than poles. This difference
appears to persist away from the strict zero and infinite coupling
limits, possibly suggesting a phase transition in large $N$ thermal 
${\mathcal N} = 4$ SYM theory as a function of the 't Hooft coupling.

\end{abstract}

\pagebreak
\setcounter{page}{1}

\tableofcontents

\section{Introduction and summary}

The AdS/CFT correspondence relates gravitational and field theories in
complementary regimes of tractability \cite{Maldacena:1997re}.
This fact is both an important success and a significant limitation of the
correspondence. On the one hand, such complementarity has allowed
insight into the dynamics of strongly coupled gauge
theories. However, if one wishes to study quantum gravity processes
involving large curvatures in the interior but
with asymptotically weak curvatures, such as evaporation of black holes,
then the appropriate dual description is a strongly coupled field
theory. Thus neither classical bulk physics nor perturbative boundary
physics is able to provide significant insight into these processes.

It is therefore an important recent observation that gravitational
physics in asymptotically AdS space with large curvatures everywhere
appears to posses a similar phase structure to gravity with
asymptotically weak curvatures
\cite{Sundborg:1999ue,Aharony:2003sx,Aharony:2005bq,
  Alvarez-Gaume:2005fv,Aharony:2004ig}.
In particular, there is a Hawking-Page phase transition
\cite{Hawking:1982dh} and a description in terms of an effective
potential with two minima and a local maximum that may be interpreted as
thermal AdS space, the large AdS black hole and the small AdS black
hole, respectively. The emerging picture suggested that perturbative
field theory could provide an inroad for studying quantum gravity
processes that were at least qualitatively similar to those of real
physical interest. Thus for instance one might hope that generalising
the work of \cite{Aharony:2003sx,Aharony:2005bq,Alvarez-Gaume:2005fv}
to consider out of equilibirum real time thermal field theory on $S^3
\times \R$ could provide a framework for studying black hole
evaporation.

The main thrust of this work is to exhibit in some detail one aspect
at least in which black holes in strongly curved AdS space are
qualitatively different to the more familiar weakly curved black
holes. One consequence of these differences will be that the
strongly curved black holes are not particularly black. Fields decay
exponentially in weakly curved AdS black hole
backgrounds \cite{Horowitz:1999jd,Ching:1995tj},
with a timescale set by the black hole temperature. In free field
theory on flat space, we see that the decay is generically only power
law whilst 
including weak interactions results in a power law times a
slow exponential decay over the plasmon damping timescale $1/(\lambda T
\ln(\lambda^{-1}))$ , 
where $\lambda \ll 1$ is the 't Hooft coupling.
This behaviour is in line with recent observations in \cite{Aharony:2005cx}, 
who noted that black holes dual to weakly coupled field theories are not
totally absorbent.

We will study the finite temperature, frequency space retarded
propagators of the scalar 
glueball operators $\Tr
F_{\mu \nu} F^{ \mu \nu}$ and $\Tr F_{\mu\nu} \widetilde F^{\mu\nu}$ at
both strong and weak 't Hooft coupling in Yang-Mills theory in
Minkowski spacetime. The ${\cal N}=4$ supersymmetric Yang-Mills (SYM)
theory at finite temperature and in flat space is realised in a
deconfined phase. In this context the operators we study are dual to 
the dilaton and axion (the RR scalar) fields, respectively, in the
bulk AdS black hole background. 
At weak coupling, a gauge
theory computation indicates that these propagators generically
contain branch cuts. At strong coupling, the same
propagators have poles and no branch cuts as found in, for example,
\cite{Son:2002sd,Starinets:2002br}. We build on previous
computations of the strong coupling behaviour to argue that these
poles persist under $\a'$ corrections without the introduction of
branch cuts. This difference in analytic structure is directly related
through a Fourier transform to the differing time decays for the fields
mentioned in the previous paragraph. Interestingly, we find that the weak
coupling frequency
space propagators bear striking similarities to 
corresponding objects computed in topological black hole backgrounds
such as the BTZ black hole.

The structure of the paper is as follows. In section 2 we compute
precisely the free Yang-Mills glueball propagator on $\R^{1,3}$, giving both
the time dependence and frequency space results. We also consider the
effects of perturbative corrections to these results, noting in
particular the resummations that are necessary due to hard thermal
loops and collisions \cite{Braaten:1989mz,Arnold:1997gh,Arnold:2002zm}. 
In section 3 we recall the prescription for computing finite
temperature correlators from bulk black hole backgrounds
\cite{Son:2002sd,Herzog:2002pc}. We
use this prescription to compute and plot the retarded propagators in
frequency space. In
section 4 we prove that at least some of the strong coupling poles
persist under $\a'$ corrections whilst branch cuts do not appear in
their vicinity. Finally, section 5 contains a
discussion in which we speculate on possible interpretations of our
results, such as the existence of a phase transition in large $N$ thermal
${\mathcal N} = 4$ SYM theory at an intermediate 't Hooft coupling,
and consequent future directions for research. Appendix A explains the
effect on the correlators of including the thermal masses and widths,
and Appendix B
contains computations of free thermal glueball correlators on $S^3
\times \R$.

\section{Thermal glueball propagators on $\R^{1,3}$}

We begin by considering $SU(N)$ ${\cal N}=4$ supersymmetric Yang-Mills
(SYM) theory at finite temperature $T$ in four dimensional Minkowski
spacetime and at small 't Hooft coupling,
\be\label{eq:coupling}
\lambda=g^2_{YM} N \ll 1.
\ee
In this section we will analyse the finite
temperature, real time
behaviour of two point correlators
of the marginal, $SO(6)$ singlet scalar glueball operators
\be\label{eq:simpleglueball}
\En = \Tr F_{\mu \nu} F^{ \mu \nu} \,,\qquad\qquad\widetilde{\cal
  G}=\Tr F_{\mu\nu} \widetilde F ^{\mu\nu}.
\ee
These are chiral operators in the ${\cal N}=4$ theory.
In the Euclidean or static description wherein the thermal field theory
is formulated on
${\mathbb R}^3 \times S^1$ these operators couple to scalar glueball
states of the three dimensional effective theory.

The temporal decay of fluctuations or correlations in a {\em weakly coupled}
Yang-Mills plasma at finite temperature will generally occur at several
distinct time scales. At short time scales set
by the inverse temperature, $t\sim 1/T$, the behaviour is completely
determined by free field theory. Intermediate and late time behaviours
are controlled by other physical scales such as the thermal gluon mass
scale $(\sqrt \lambda  T)^{-1}$, the gluon damping rate
$(\lambda T)^{-1}$, etc. \cite{Arnold:1997gh}.

\subsection{Short times or the free theory}

For short time scales $t\sim T^{-1}$, and weak coupling, thermal
correlators can be approximated by straightforward diagrammatic
perturbation theory. Thus, at leading order we only need the quadratic terms
in the Yang-Mills action and the glueball operator. Furthermore, we
can focus attention solely on the transverse, physical degrees of
freedom, since the longitudinal mode only mediates an instantaneous Coulomb
interaction and does not affect the dynamics at any finite time $t>0$.
We also therefore don't need to worry about ghost fields.
The time ordered two point function for the transverse gauge
field is
\be
\langle A^{a}_i(t,{\bf p}) A^{b}_j(0,{\bf p}) \rangle =
D^{>(ab)}_{ij}(t)\;\Theta(t) +  D^{<(ab)}_{ij}(t)\;\Theta(-t) \,
\ee
where ${\bf p}$ is the spatial momentum and \cite{Niemi:1983nf,Kobes:1984vb}
\be
\label{eq:freept}
D^{>(ab)}_{ij}(t) =  D^{<(ab)}_{ij}(-t)=i \left(\d_{i
    j} - \frac{p_i p_j}{|p|^2} \right) \frac{\d^{a b}}{2 E}
\left[(1+n(E)) e^{-i E t} + n(E) e^{i E t} \right] \,.
\ee
Here $E = p \equiv |{\bf p}|$ and $n(E)$ is the Bose-Einstein distribution
\be
n(E) = \frac{1}{e^{\b E} - 1}\;,\qquad\beta={1\over T}\,.
\ee
The glueball two point function with external momentum ${\bf p}$
is then straightforwardly worked out using Wick contractions to be
\bea\label{eq:ptfn}
\lefteqn{ \langle \En^p(t) \En^p(0) \rangle =
\langle {\widetilde \En}^p(t) {\widetilde \En}^p(0) \rangle
\equiv \int {d^3 x} \;e^{-i {\bf
     p\cdot {\bf x}}}\langle{\cal G}(t, {\bf x}) {\cal G}(0)\rangle } 
\nonumber\\
& = & -\frac{N^2}{2 \pi^3} \int \frac{d^3 p'}{|{\bf p}'| |{\bf p}-{\bf
     p}'|} \left[
   \left({\bf p}'^2 ({\bf p}-{\bf p}')^2
+ [{\bf p}' \cdot ({\bf p}-{\bf p}') ]^2 \right) F(|{\bf p}'|,t)
F(|{\bf p}- {\bf p}'|,t) \nonumber \right. \\
& & \left. + 2 \, {\bf p}' \cdot ({\bf p} - {\bf p}')  \dot{F}(|{\bf p}'|,t)
\dot{F}(|{\bf p}-{\bf p}'|,t) \right] \,,
\eea
where
\be
F(E,t) = (n(E)+1) e^{-i E t} + n(E) e^{i E t} \,.
\ee
The $\En$ and ${\widetilde \En}$ glueballs have the same free theory
thermal two point functions. We will write $\En$ from now on for
notational convenience.

We first consider the homogeneous mode with no external momentum,
${\bf p}=0$. The relaxation of long-wavelength fluctuations
or `soft' modes with $p \ll T^{-1}$ can be non-trivially altered at late
times by interactions, as we discuss below.
Nonetheless in the limit of zero coupling, or
for suitably short time scales in the interacting theory,
the correlator at zero momentum, with ${\bf p}=0$ in (\ref{eq:ptfn}), becomes
\be\label{eq:homint}
\langle \En^0(\t) \En^0(0) \rangle = - \frac{8 N^2}{\pi^2} T^5 
\int_0^{\infty}
dx\, x^4 \frac{e^{2x} e^{-i 2 x \t} + e^{i 2 x \t}}{(e^x-1)^2}\,,
\ee
where we have introduced the dimensionless time $\t = t T$ and the
variable of integration $x = |{\bf {p}}'|/T$.
The integral may be evaluated to give
\be
\langle \En^0(\t) \En^0(0) \rangle = \frac{N^2}{2 \pi} T^5 \left[8
  \frac{d^3}{d\t^3} + (2\t + i) \frac{d^4}{d\t^4} \right] \coth(2\pi
\t) \,.
\ee
As $\t \to 0$ there is a $1/\t^5$ divergence, which is the zero
temperature power law behaviour. Importantly, this expression exhibits
an exponential decay on time scales $\tau\sim 1$ or $t\sim T^{-1}$.

Using the above result we obtain the retarded propagator which is of
particular physical interest and is given by
\be\label{eq:homogret}
\langle \En^0(t) \En^0(0) \rangle_R =
-i \Theta(t)\langle[\En^0(t),\En^0(0)]\rangle = \frac{N^2}{\pi} T 
\;\Theta(t) \frac{d^4}{dt^4} \coth(2\pi T t) \,.
\ee
The retarded propagators naturally occur in the framework of
linear response  theory wherein one probes the plasma with a weak
disturbance and studies the response of the medium in a linearised
approximation.

In the AdS/CFT correspondence,
retarded thermal propagators of the boundary field theory are known
to encode physical properties of AdS$_5$ black hole
backgrounds, at least in the classical gravity approximation which is
dual to strongly coupled field theory. Specifically, poles of
retarded propagators in the frequency domain correspond to
quasinormal frequencies of AdS black holes
\cite{Birmingham:2001pj,Son:2002sd}. The scalar glueball operators
$\En$ and ${\tilde \En}$
are dual to the dilaton and axion in the bulk, respectively.
The associated retarded correlators
{\em at strong coupling} encode quasinormal frequencies of massless
minimally coupled scalar perturbations about the AdS$_5$ black hole 
background.
We will return to the gravitational computation of correlators in
later sections of this paper.

A characteristic feature of scalar perturbations about AdS black hole
spacetimes is that they always
decay exponentially \cite{Horowitz:1999jd,Ching:1995tj}. The
exponential decay (\ref{eq:homogret}) that we have encountered in the
free theory might seem to be consistent with this feature, although it
is in  
a
regime where the dual geometry is very strongly curved. However, the
exponential decay of quasinormal perturbations in AdS black hole
spacetimes represents the decay of perturbations in the boundary
thermal theory and its consequent approach towards equilibrium. The
exponential decay that we see for the homogeneous mode
in the free theory is not, strictly
speaking, an approach to equilibrium since scattering and collisional
processes are absent in this limit. The decaying transient that we see
here arises from interference effects between the different
modes in (\ref{eq:homint}). In fact we will see shortly that all the 
inhomogeneous
modes of (\ref{eq:ptfn}) in the free theory have oscillating power law
decays instead, a fact which will have a simple physical explanantion.

The qualitative similarities and dissimilarities between the field
theory at weak and strong 't Hooft couplings are naturally captured by
the analytic structure of the retarded propagator in frequency
space. Fourier transforming equation (\ref{eq:homogret}) we find
\be\label{eq:fourier}
G^0_R(\omega) = - \frac{N^2}{2\pi^2}\;\w^4
\left[\psi\left(\frac{-i\;\w}{4\pi T} \right) - {1\over 3}
\left(\frac{2 \pi T}{ \w}\right)^2-
  {2\over 15}\left(\frac{2 \pi T}{ \w}\right)^4 +
  i \frac{2 \pi T}{\w} \right] \,,
\ee
where $\psi(z) = \Gamma'(z)/\Gamma(z)$, the logarithmic derivative
of the Gamma function.
In the zero temperature limit, using Stirling's approximation for
$\Gamma(z)$ at large $z$, we have
\be
G^0_R(\omega)\approx -{N^2\over 4 \pi^2}\;\omega^4 (\log \frac{\w^2}{\mu^2}
-i \pi\;\text{sgn}\;\omega) + \cdots \quad \text{as}
\quad {\omega\over T} \to \infty \,,
\ee
which is the known zero temperature result (see {\it
  e.g.}\cite{Son:2002sd}).

The Fourier transformed homogeneous correlator (\ref{eq:fourier}) is
analytic in the upper half plane $\Im\;\omega>0$. This is always true
for retarded propagators with our conventions. On the other hand the
function has an infinite set of simple poles along the lower
imaginary axis at
\be
\omega_n= -4 \pi i T n\;,\qquad n=1,2\ldots
\ee
These poles, illustrated in figure 1,
lead to the exponential decay in time of (\ref{eq:homogret}).
This analytic structure at zero 't
Hooft coupling will of course be modified by a small nonzero value of
$\lambda$ and we will discuss the modifications in more detail
below. Perhaps unsurprisingly, the analytic structure at $\lambda=0$
bears little resemblance to that at $\lambda\to\infty$ wherein the
retarded correlator has an infinite number of simple poles at
locations that at large $n$ tend to $\;\omega_0^\pm \pm 2 \pi T \,n \,(1\mp 
i)$ where
$\omega_0^\pm\approx \pm 1.2139-0.7775 i$
\cite{Son:2002sd}. The strong coupling poles are sketched in figure 2.
The strong-weak coupling discrepancy will be substantially more severe
for the inhomogeneous modes.

\begin{figure}[h]
\begin{center}
\epsfig{file=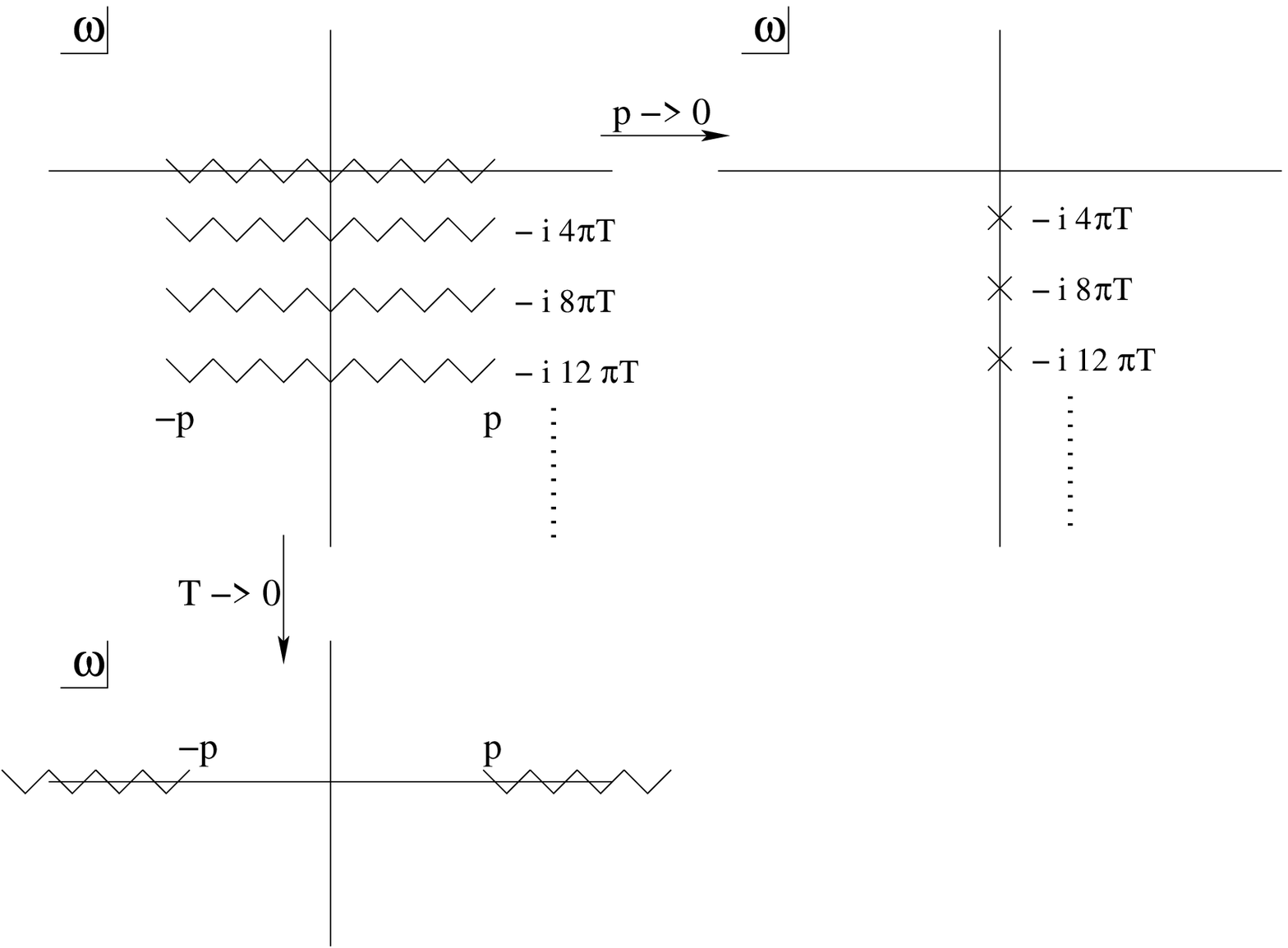,width=12cm}
\end{center}

\noindent {\bf Figure 1:} Poles and branch cuts of the free theory
propagator $G_R^p(\w)$.
Top left is the full result with momentum $p$. Top right
is the homogeneous mode with $p=0$. Bottom left is the
zero temperature limit with momentum $p$.
\end{figure}

It is interesting to note that the frequency space propagator
(\ref{eq:fourier}) is, up to a dimension dependent phase space
factor, precisely the function encountered for the propagator
at zero spatial momentum
in the general BTZ black hole background dual to a two dimensional
CFT \cite{Son:2002sd}. Drawing further on this analogy,
perhaps this suggests that
like the BTZ black hole, the strongly curved five dimensional
background is a kind of topological black hole (c.f. higher
dimensional AdS space with global identifications
in for instance \cite{Mann:1997iz, Birmingham:1998nr,
  Banados:1998dc}). From the
free field theory perspective, the similarity to a two dimensional
CFT and the BTZ black hole arises simply because free field
correlators are completely
determined by the scaling dimensions of the operators in question and
the effect of finite temperature is to require an infinite
sum over images that renders the correlators periodic in imaginary
time.

The inhomogeneous contribution to the propagator
(\ref{eq:ptfn}), with spatial momentum ${\bf p}$, may also be calculated.
Introducing the dimensionless momentum
${\bf k} = {\bf p}/T$, the retarded propagator is given in terms of
the integral
\bea\label{eq:double}
\langle \En^p(\t) \En^p(0) \rangle_R & = & \frac{- N^2 \Theta(\t)}{4
  \pi^2 }{T^5\over k}
\int_k^{\infty} du \int_{-k}^k dv \left[ \frac{(u^2-k^2)^2 (1-
    e^u)}{(e^{(u+v)/2}-1)
  (e^{(u-v)/2}-1)}
    \sin (u \t) \right. \nonumber \\
& - & \left. \frac{(k^2-v^2)^2 e^{u/2} (e^{v/2} - e^{-v/2})}{(e^{(u+v)/2}-1)
  (e^{(u-v)/2}-1)} \sin(v \t) \right] \,.
\eea
The double integral (\ref{eq:double}) may be evaluated, partly using
contour integration, to give
\be\label{eq:inhomo}
\langle \En^p(t) \En^p(0) \rangle_R = \frac{N^2 \Theta(t)}{\pi^2}
        {T\over p}\left(p^2 + \frac{d^2}{dt^2} \right)^2 \left[\frac{\sin
  p t}{t} \pi \coth (2\pi t T) - \frac{\sin p t - p t \cos p t}{2 t^2 T} 
\right].
\ee
As $t \to \infty$ this propagator decays as an oscillatory
power law rather than exponentially. The physical reason for this is
that these modes carry a conserved momentum which in the free theory
cannot dissipate into other modes. In the strongly coupled theory
($\lambda \rightarrow \infty$),
dual to the classical black hole, the momentum can quickly dissipate
into other modes and so one expects an exponential decay in this case,
as indeed occurs.
A small but nonzero $\lambda$ will also modify this
power law decay at intermediate and late times as we discuss below.

The propagator (\ref{eq:inhomo}) may be Fourier transformed to
frequency space to yield
\bea\label{eq:fourierz}
G^p_R(\w) & = & - \frac{N^2}{\pi^2}(p^2-\w^2)^2
  \left[\frac{1}{2} + \left(i\frac{\pi T}{2 p} -
\frac{\w}{4p} \right) \log \frac{\w + p}{\w - p} + i\frac{\pi T}{p} \log
    \frac{\G\left(\frac{-i(\w+p)}{4\pi
        T}\right)}{\G\left(\frac{-i(\w-p)}{4\pi T}\right)}\right]
  \nonumber \\
&  & + \frac{N^2}{\pi^2}\left[\frac{2 \pi^2 T^2}{3} (\w^2-p^2) +
    \frac{16 \pi^4 T^4}{15} +
  \frac{p^2}{6} \left(\frac{7 p^2}{5} - \w^2\right) \right]\,.
\eea
It is easily seen that (\ref{eq:fourierz}) reduces to
(\ref{eq:fourier}) in the vanishing momentum $p \to 0$
limit. In the zero temperature limit we also recover the wellknown
Lorentz invariant result, providing a quick check of (\ref{eq:fourierz})
\be
G^p_R(\w) = \frac{- N^2}{4\pi^2} (p^2-\w^2)^2 \log
\frac{p^2-\w^2}{\mu^2} \,.
\ee

The central point to note here is that the frequency space result
(\ref{eq:fourierz}) has only branch cuts and no poles, as
illustrated in figure 1. The infinite set of simple poles of the
zero momentum correlator (\ref{eq:fourier}) have smeared out into
branch cuts stretching between $\w=-4 \pi i T n - p$ and $\w=-4 \pi i
T n + p$ with $n=1,2, \ldots$. In the time domain, these branch cuts centred
around points on the negative imaginary axis translate into
exponential falloff accompanied by power law and oscillatory corrections. In
addition to these, there is also a branch cut on the real axis between
$\w=-p$ and $\w=+p$ and it is this branch cut which is responsible for the 
leading
oscillatory power law behaviour in time in (\ref{eq:inhomo}).

These results may be contrasted with the analytic structure emerging from 
dual
gravitational computations that describe the strongly
coupled regime, e.g. \cite{Starinets:2002br, Son:2002sd}. The strong
coupling results are sketched in figure 2, displaying an infinite set of
simple poles corresponding to the quasinormal frequencies of the AdS$_5$ 
black
hole background. These asymptote to $\omega_0^\pm(p)\pm 2\pi T n (1\mp i)$.
Note that both the real and imaginary parts of the
quasinormal frequencies grow linearly with $n=1,2,\ldots$.

\begin{figure}[h]
\begin{center}
\epsfig{file=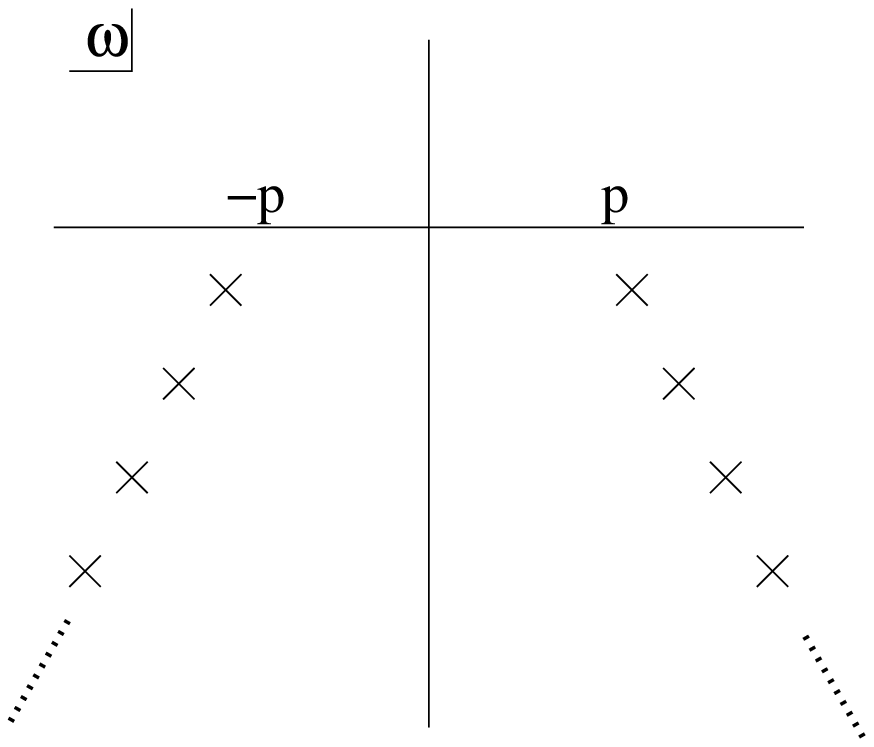,width=5cm}
\end{center}

\begin{center}{\bf Figure 2:} Poles in the strongly coupled
propagator \cite{Starinets:2002br}.\end{center}
\end{figure}

Thus we can see that correlators in the $\lambda=0$ theory exhibit
features that are qualitatively different to those of the
$\lambda\to\infty$ theory. As we
will argue in the following section, moving slightly away from
the $\lambda=0$ point does not
change this qualitative picture drastically, although it does affect
the late time behaviour of the correlator (\ref{eq:inhomo}) causing it
to decay exponentially. While branch cuts in frequency space induce
power laws in the time domain, the endpoints of the branch cuts
dictate the exponential decay constants and/or oscillatory
behaviour. For fixed $p$ the real parts of the branch
point locations of (\ref{eq:fourierz}) are fixed while the imaginary part
increases with $n$. Following our previous comments concerning
parallels with the BTZ black hole, it is tantalising to note that
quasinormal frequencies of topological AdS black holes appear to have a 
similar
property, namely that for fixed spatial momentum their real parts are
constant while the imaginary parts take an infinite set of discrete
values \cite{Aros:2002te}.

Finally, we have already noted that the operator $\widetilde {\cal G}({\bf x}, 
t)$ in (\ref{eq:simpleglueball}) has exactly the same two point function
as ${\cal G}$ in free field theory. This operator has the quantum
numbers of the $0^{-+}$ scalar
glueball, and at strong coupling it is dual to the axion (the RR scalar) in
supergravity. We will see shortly that this implies that also at
strong 't Hooft coupling the
operators ${\cal G}$ and $\widetilde {\cal G}$ share the same
correlators. At finite values of $\lambda$ however, we expect them to
differ quantitatively.

\subsection{Late times and weak coupling $\lambda\neq 0$}

We now turn to the effects of nonzero 't Hooft coupling $\lambda$ on
the late time behaviour of the correlators above. In the interacting
${\cal N}=4$ SYM theory, the correlators of interest receive higher
order contributions from gluon self-interactions and from scalar and
fermion matter fields in the adjoint 
representation of the gauge group. Since the thermal bath breaks SUSY
and conformal invariance, in what follows  it is sufficient to
focus on the known general qualitative aspects of time dependent
correlators in a weakly  
interacting Yang-Mills plasma in the deconfined phase. In the weakly
coupled plasma there is a hierarchy of energy scales, $(T,
\sqrt\lambda T, \lambda T,\ldots)$ corresponding to distinct physical phenomena
that dictate the temporal behaviour of correlators over the associated
characteristic time scales.

In order to go
beyond the leading perturbative result it is necessary to implement the finite
temperature improvement of resumming the hard thermal loops
\cite{Braaten:1989mz}. Essentially this means that
%when the momenta on
for the internal lines with momenta
${\bf p}'$ or ${\bf p}-{\bf p}'$ 
in the above one loop
calculation,
%are $\lesssim \sqrt\lambda T$, then 
we should include the thermal self-energy corrections.  
Specifically, for the hard momenta of order $T$ which dominate the
graph, this self-energy correction leads to 
an effective thermal mass $m_T\sim \sqrt\lambda T$. In addition there
is also a higher order effect leading to a 
thermal width, or more 
precisely the plasmon damping rate for hard gluons, $\Gamma\sim
\lambda T\ln(\lambda^{-1})$. This resummation has a
significant effect on homogeneous correlators or those with `soft'
external momentum $p\lesssim m_T\sim\sqrt\lambda T$. For `hard' modes, with
external momentum $p\gtrsim T$, the effect of the hard thermal loop
resummation remains perturbatively small and does not significantly
alter the early time oscillatory power law behaviour found in
(\ref{eq:inhomo}).

First consider the effect of incorporating thermal masses for internal
lines. This will shift the
threshold branch points at $\omega=\pm p$ to $\omega=\pm
\sqrt{p^2+4 m_T^2}$. The thermal mass thus changes the analytic
structure of the homogeneous correlator with $p=0$ (\ref{eq:fourier}) by
introducing a branch cut
between $\omega=+2 m_T$ and $\omega=-2 m_T$ which leads to
oscillatory power law decays in time. This change in the
analytic structure can be inferred from (\ref{eq:ptfn}) by setting
$p=0$ and using the Bose-Einstein distribution for a massive particle.
We perform this computation in Appendix A.

It is easy to see, following the example in
Appendix B of \cite{Arnold:1997gh}, that this inclusion of the thermal
mass
leads to a power law decay with oscillatory behaviour for time
scales $t>(\sqrt\lambda T)^{-1}$, where the oscillations are controlled
by the thermal mass $m_T$. Hence the short time exponential decay of
the homogeneous correlator (\ref{eq:homogret}) turns into an
oscillatory power law on time scales of order $(\sqrt\lambda
T)^{-1}$. This power law behavior kicks in when the free field
correlator has decayed to $\sqrt\lambda$ times its value at $t=0$.

However, including only thermal masses is, strictly speaking, a
collisionless approximation from the point of view of quasiparticles
in the plasma although it includes quantum field theory interactions.
What happens eventually to all correlators at late times is actually
determined by scattering and collisions of quasiparticle
states. A detailed quantitative treatment of this question is beyond
the scope of this work. However, much can be learned by following the
arguments in \cite{Arnold:1997gh}. Incorporating the thermal
widths of intermediate states, the imaginary part $\Gamma$ of the self
energy, 
makes internal propagators decay on time scales of order $(\lambda
T  \ln(\lambda^{-1}))^{-1}$ and hence the correlators themselves decay
exponentially on
these time scales. This is a somewhat subtle point as discussed in
\cite{Arnold:1997gh}. The effect of incorporating the width, or 
plasmon damping rate, can in certain cases actually cancel out
between different diagrams in resummed  perturbation theory so that
the exponential decay time scale is longer and is dictated by
different physics. In particular, this occurs for correlators
involving conserved currents wherein the time scale for exponential
decay is controlled by that of large angle
scattering $\sim (\lambda^2 T \ln(\lambda^{-1}))^{-1}$ and can be seen using
an effective kinetic theory 
description \cite{Arnold:2002zm}\footnote{Correlators for 
conserved currents in a finite $N$ theory 
eventually decay via hydrodynamic power law tails 
which are not visible in the large $N$ limit
\cite{Kovtun:2003vj}. These $1/N$ suppressed 
power law tails correspond to non-analyticities in the frequency space
correlators at small $\w$.}. Since the scalar glueball
operators are not associated to any conserved currents or charges, we
expect them to decay on time scales set by plasmon damping which is
also referred to in \cite{Arnold:1997gh}
as the time scale for `any-angle scattering'.  

The picture that thus emerges when finite coupling
$\lambda$ is considered is that the analytic
structure of the glueball two point function in frequency space is
qualitatively similar to figure 1, with an extra branch cut about the origin
along the real axis in the homogeneous case. But
this description cannot account for the late time exponential decay
argued for above since the branch cut along the real axis in the frequency
plane seen in the lowest order approximation would lead to oscillatory
power laws at late times. This discrepancy suggests the following possibility:
\begin{itemize}
\item[{\bf (1)}] When all relevant perturbative corrections
are accounted for, the branch points on
the real frequency axis are shifted downwards into the lower half plane
by an amount $\sim i \lambda T
\ln(\lambda^{-1})$. A na\"{\i}ve calculation illustrating this effect is given
in Appendix A.
\end{itemize}
This analytic structure for frequency space propagators would be
consistent with power law tails at intermediate times accompanied by
exponential decay times power laws at late times. A confimation of
this possibility would require a detailed consideration of the
correlators with full self-energy resummations for the internal lines.
Thus we cannot rule out the following further possibility: 
\begin{itemize}
\item[{\bf (2)}] The branch cut seen in the lowest order approximation from
$\w=\pm 2m_T$ for the homogeneous correlator, and from
$\w\approx \pm p$ for the inhomogeneous cases, might
split into a series of poles with perturbatively small separations,
which then move off the real axis upon inclusion
of all perturbative corrections in the large $N$ limit of the ${\cal
  N}=4$ theory.
\end{itemize}
Whilst we have no evidence in favour of this second
possibility, it would reproduce the expected intermediate and late
time behaviour in the weakly coupled plasma, and would also allow a
smooth interpolation to the strongly coupled gravity dual picture
discussed in the following section. 

Regarding the second possibility above, it is not clear
how poles might emerge from cuts at weak
coupling. Indeed, perturbative results for real time response functions in
weakly coupled plasmas in the 
deconfined phase generically give rise to branch cuts in correlators
from zero frequency  (see e.g.\cite{Pisarski:1987wc,Boyanovsky:2004dj}).

Our discussion in this section has consequences for bulk physics.
The strong coupling exponential decay on timescales $t \sim T^{-1}$ is
naturally interpreted gravitationally as being due to matter falling
into the black hole. The much slower decays we have just described at
weak coupling therefore suggest that black holes in highly curved AdS
backgrounds are less absorbent. This phenomenon is consistent with the
recent comments in \cite{Aharony:2005cx}.

\pagebreak

\section{Strong coupling correlators from gravity}

A prescription for computing thermal two point correlators in strongly 
coupled
${\mathcal{N}}=4$ SYM theory was developed in
\cite{Son:2002sd,Herzog:2002pc}. This method has allowed the
computation of poles in the Minkowski space retarded thermal propagator
via bulk quasinormal modes
\cite{Starinets:2002br,Nunez:2003eq}, with particular
success in the hydrodynamic limit \cite{Policastro:2002se}.

In this section we review the bulk prescription and use it to produce plots
of the thermal propagator as a function of frequency $\w$. These weak
curvature results describe the strongly coupled SYM theory.
We will compare with plots for the free
theory correlators following from our results in the previous section,
highlighting various qualitative differences.
In the following section we will show that the strong coupling correlators
do not have branch cuts, as was implicitly anticipated in
\cite{Starinets:2002br,Nunez:2003eq} and thus confirm figure 2. Our
results on the absence of branch cuts will extend to include
$\a'$ corrections.

For notational clarity when discussing gravitational computations, we
shall absorb $T$ into the frequency: $\w/T \to \w$.

As we wish to compare with field theory on $\R^{1,3}$, the bulk
is given by the near horizon geometry of a stack of non-extremal D3
branes
\be\label{eq:blackbrane}
ds^2 = \frac{R^2}{z^2} \left[-f(z) dt^2 + d{\bf x}^2 +
\frac{dz^2}{f(z)} \right] + R^2 d\Omega^2_5 \,,
\ee
where as usual $d\Omega^2_5$ is the round metric on a five sphere, $R$
sets the AdS scale and
\be
f(z) = 1 - \frac{z^4}{z_H^4} \,,
\ee
with the horizon $z_H = \b / \pi$.

The prescription may now be stated as \cite{Son:2002sd,Herzog:2002pc}
\be\label{eq:prescription}
G_R(\w) = \left. K \sqrt{-g} g^{zz} \phi_\w(z) \pa_z \phi_\w^*(z)
\right|_{z_B \to 0} \,,
\ee
where $K$ is a normalisation constant. The normalisation is known from
the bulk action \cite{Son:2002sd} or alternatively can be fixed
by the normalisation of the zero temperature behaviour $\w^4 \log \w$
as $\w \to \infty$. The zero temperature result will agree with the
boundary free field theory computation because the $\Tr F^2$ operator
is BPS. The bulk modes $\phi(z)$ are
evaluated near the boundary $z_B \ll R$ where they are required to
satisfy $\phi(z_B) = 1$. They further satisfy incoming wave boundary
conditions at the horizon
\be\label{eq:incoming}
\phi_\w(z) \sim (z_H - z)^{-i \w/4\pi} \quad \text{as} \quad z \to z_H \,.
\ee
As usual in computing boundary correlators using the AdS/CFT
correspodnence, the expression (\ref{eq:prescription}) will contain
volume divergences that correspond to UV divergences in the dual
theory. In order to meaningfully compare bulk and boundary results, we
need to fix a regularisation scheme. We do this as follows: Firstly in
(\ref{eq:prescription}) we minimally subtract off all the divergent
terms as $z_B \to 0$. This gives a finite propagator whose only scheme
dependence is due to the ambiguous subtration of a logarithmic
divergence $\sim \w^4 \log z_B/\mu$ . In order
to extract scheme independent information we then
subtract off from the propagator the terms that diverge like $\w^4 \ln \w$ 
and $\w^4$ as $\w \to \infty$.
This subtraction corresponds to removing the zero
temperature result. We will perform the same subtractions on our
propagators from the previous section.

The bulk field dual to the glueball operator $\Tr F^2$ is the
dilaton, whilst the field dual to $\Tr F {\widetilde F}$ is the axion.
Both of these fields are minimally coupled scalars and therefore have a
mode equation in the black brane background (\ref{eq:blackbrane})
given by \cite{Son:2002sd,Starinets:2002br}
\be\label{eq:bulkeqn}
\phi'' - \frac{1+u^2}{u(1-u^2)} \phi' + \frac{w^2/4\pi^2}{u(1-u^2)^2} \phi -
\frac{p^2/4\pi^2}{u(1-u^2)} \phi = 0 \,,
\ee
where the coordinate $u = z^2/z_H^2$. By solving this equation
numerically and varying $\w$ and $q$, we can plot the propagator
(\ref{eq:prescription}). The real part of the propagator has a
more interesting structure than the imaginary part,
so we plot this. Furthermore, the real
part of $G_R(\w)$ is symmetric on the real axis, so we restrict
plots to $\w \geq 0$. It is curious that these two operators have
exactly the same thermal propagator both at zero and at strong
coupling.

Figures 3 and 4 show the propagator for the homogeneous mode
as computed in free field theory (\ref{eq:fourier}) and from the bulk using
(\ref{eq:prescription}), respectively. As discussed above, the zero
temperature result has been subtracted to ensure scheme
independence
\be
\Delta G_R^p(\w) = G_R^p(\w) - \# (p^2-\w^2)^2 \log \left(p^2 - \w^2
\right) \,.
\ee
Although the plots show some similarity, and curiously both lack an
$\w^2$ growth as $\w \to \infty$, they remain fairly distinct,
consistent with their differing analytic structure off the real
axis.

\begin{figure}[h]
\begin{center}
\epsfig{file=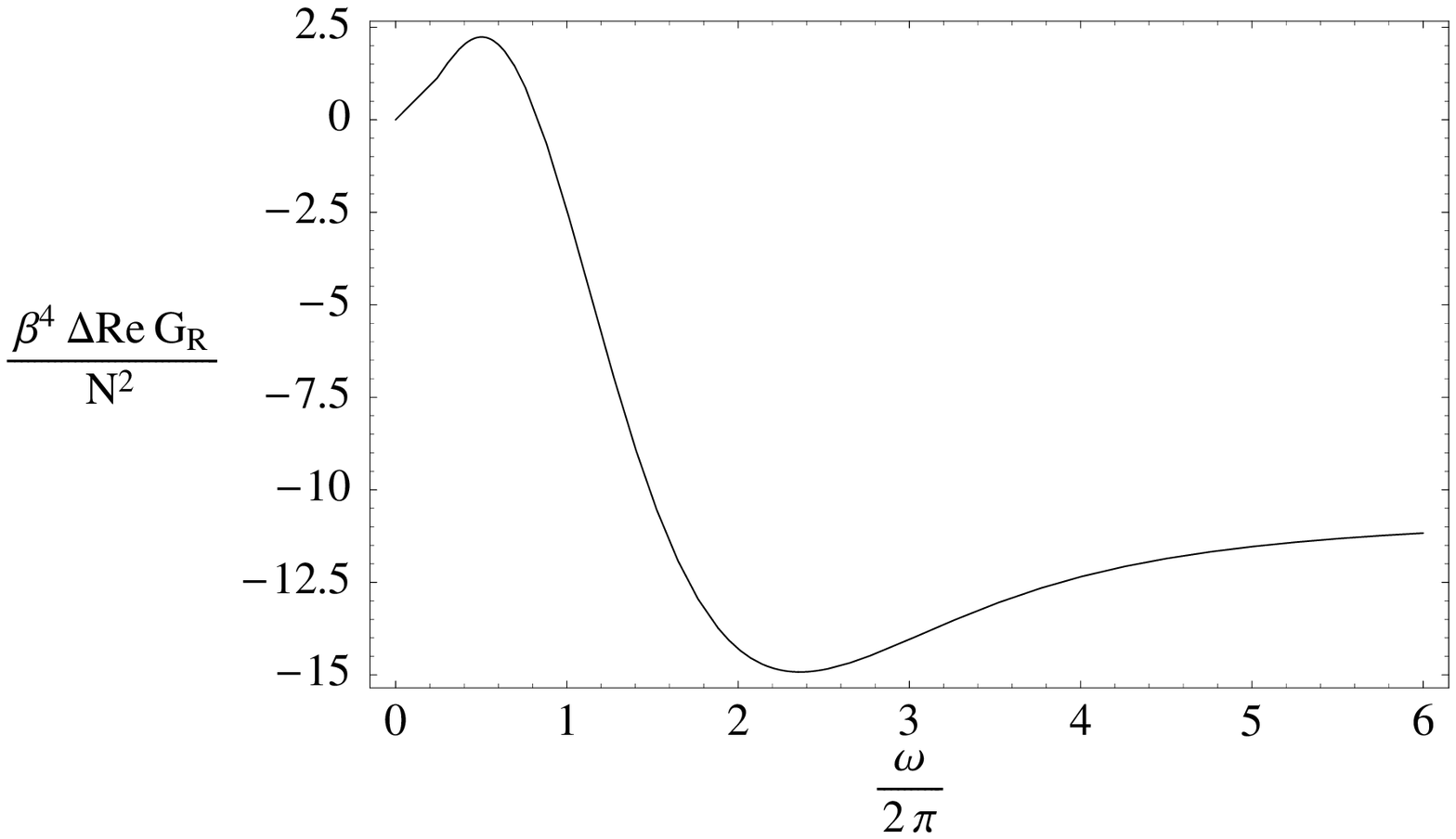,width=10cm}
\end{center}

\begin{center} {\bf Figure 3:} The real part of $\Delta G_R^0(\w)$
as computed in free field theory.\end{center}
\end{figure}

\begin{figure}[h!]
\begin{center}
\epsfig{file=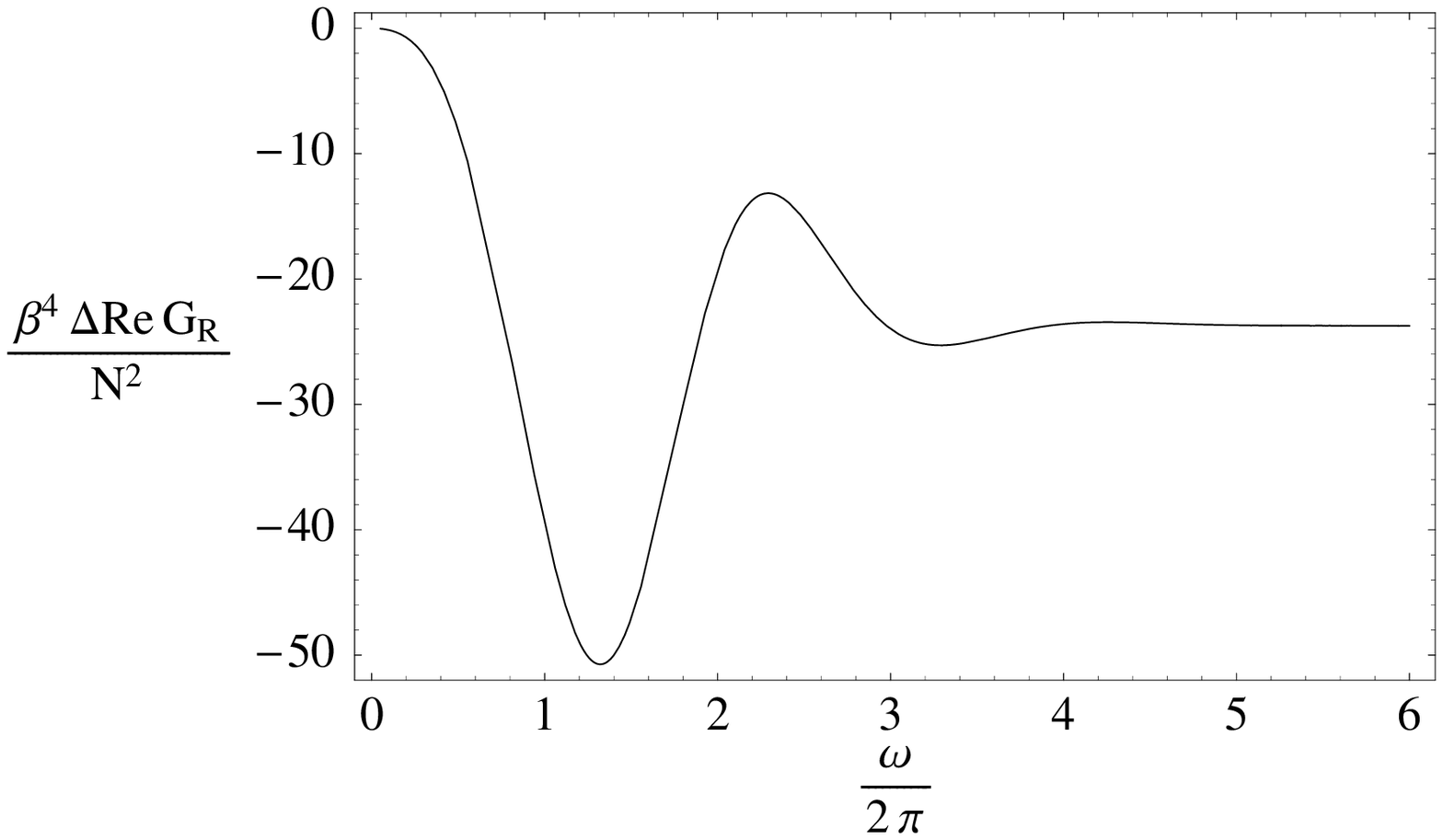,width=10cm}
\end{center}

\begin{center} {\bf Figure 4:} The real part of $\Delta G_R^0(\w)$
as computed from the bulk supergravity.\end{center}
\end{figure}

Figures 5 and 6 show the effect of including a momentum $p$ in the
propagator in free field theory and in the bulk, respetively. In
this case we have plotted for both positive and negative $\w$ to
emphasise the features of the plots.

\begin{figure}[h]
\begin{center}
\epsfig{file=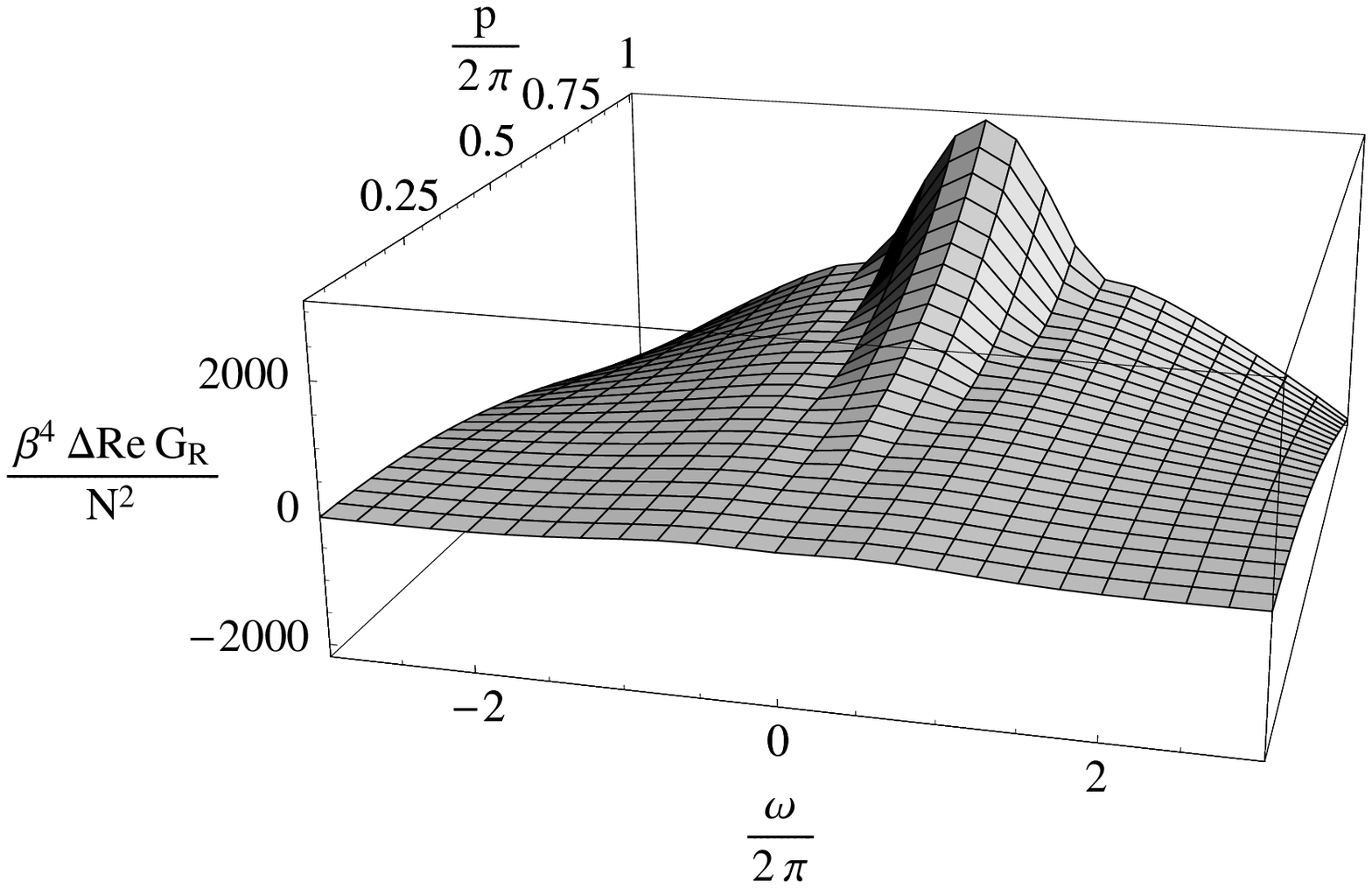,width=11cm}
\end{center}

\begin{center} {\bf Figure 5:} The real part of $\Delta G_R^p(\w)$
as computed in free field theory.\end{center}
\end{figure}

\begin{figure}[h!]
\begin{center}
\epsfig{file=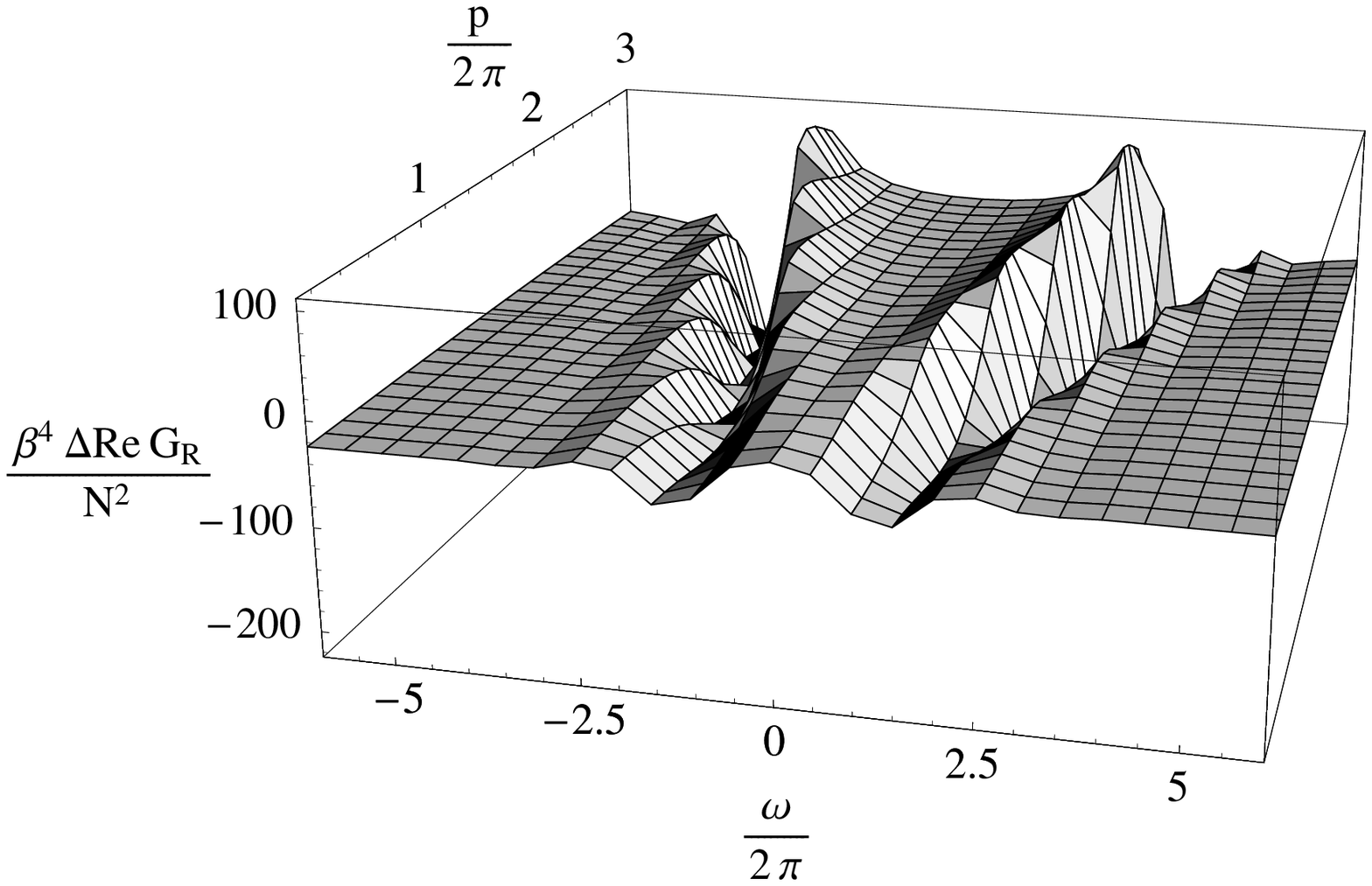,width=11cm}
\end{center}

\begin{center} {\bf Figure 6:} The real part of $\Delta G_R^p(\w)$
as computed from the bulk supergravity.\end{center}
\end{figure}

The main features in the free theory plot of figure 5
are a positive $p^4$ growth with momentum along the $\w=0$
axis which is apparent in (\ref{eq:fourierz}) and a negative $\w^2$ growth
which becomes more significant at larger $p$. The transition between
these behaviours occurs around $\w \sim \pm p$ where the branch points
are located.

In the bulk computation shown in figure 6 we see that the peak and trough 
present
already in the homogeneous case, figure 4, become more pronounced
once $p$ is included and move linearly like $\w \sim \pm p$ as $p$
increases. This is consistent
with these peaks being resonances of the nearest quasinormal pole to
the real $\w$-axis, see figure 2. It was shown in \cite{Starinets:2002br} 
that as
$p$ is increased the lowest poles move closer to the real axis and
tend towards $\w = \pm p$.

One can numerically check that there is no branch point at, for
instance, $\w = \pm p$ in the bulk
results by computing the propagator along the circles
\be
\w = \pm p + e^{i \theta} \,, \qquad \theta \in \left[0, 2 \pi \right] \,,
\ee
and verifying that both real and imaginary parts of the propagator are
continuous. In the following section we will prove the absence of
branch points in some regions of the complex $\w$ plane. This argument
will then extend to include $\a'$ corrections.

\section{Persistence of poles under $\a'$ corrections}

To move away from the strict strong coupling limit, we need to
consider string theory $\a'$ corrections to the gravity background
and to the equations of motion for the fluctuations. We will be
able to show rigorously that at least some of the poles in the
retarded glueball propagator that are found in the strong coupling
limit persist under $\a'$ corrections without becoming branch
points. The argument does not rely on the precise form of the
$\a'$ corrections, which is fortunate as even the leading order
corrections to the type IIB string theory action are not known
completely and hence precise calculations using such corrections
are not possible in our context, despite some assumptions to the
contrary in the literature.

The computation of the retarded glueball propagator from the bulk
follows the same logic as previously. One takes the corrected
background and considers linearised fluctuations of the dilaton or
axion about this background. The fluctuation equations will be higher
order and will therefore have more than two independent solutions.
However, only two of these solutions will deform into the
uncorrected solutions in the $\a' \to 0$ limit. These are the
modes dual to the glueball operator. The other modes are stringy
degrees of freedom arising in the $\a'$ corrected action, these
are dual to other boundary operators.

The boundary correlator is then calculated as usual by considering
the on shell corrected bulk action as a function of the boundary field
values. As we tend towards the boundary, $u \to 0$, the leading
$u$ dependence of the field must be the same as in the uncorrected
case, otherwise the $\a'$ corrections would not be subleading and
the perturbation expansion would not be reliable. Thus we have
from (\ref{eq:bulkeqn})
\be\label{eq:newasymptotics}
\phi_\w(u) = A(\w) \left( 1 + \cdots \right) + B(\w)
\left( u^2 + \cdots \right) \qquad \text{as} \quad u \to 0 \,.
\ee
The coefficients $A(\w)$ and $B(\w)$ are fixed by the incoming
wave boundary condition at the horizon (\ref{eq:incoming}). The
fact that the $u$ dependence near the boundary is as in the
uncorrected case, although the $\w$ dependence of
(\ref{eq:newasymptotics}) is of course different, implies that the
$G_R(\w)$ correlator has the same structure as previously in
(\ref{eq:prescription})
\be\label{eq:seepole}
G_R(\w) \sim \left. \frac{\phi_\w \pa_u \phi^*_\w}{\phi_\w \phi^*_\w}
\right|_{u = u_B \to 0} \sim \frac{B(\w)}{A(\w)} \,.
\ee
This expression should be multiplied by functions with $u$
dependence, and the evaluation as $u \to 0$ will require
regularisation. By dividing by $\phi_\w \phi^*_\w$
we have explicitly imposed the
additional normalisation $\phi(u_B)=1$ at the boundary cutoff. The
expectation that the result reduces to a $\phi_\w \pa_u
\phi^*_\w$ term is realised explicitly in
\cite{Buchel:2004di} who consider a partial set of $\a'$
corrections to the Yang-Mills shear viscosity.

The expression (\ref{eq:seepole}) was a basic connection
between bulk quasinormal modes and poles in the dual propagators
observed in \cite{Son:2002sd}. A quasinormal mode in asymptotically
AdS space satisfies incoming boundary conditions at the horizon and is
normalisable as $u \to 0$. This will occur at frequencies where
$A(\w)=0$, which hence correspond to poles in $G_R(\w)$. The
corrected fluctuation equation will have quasinormal
modes and hence dual poles that are simply deformations of the
strict strong coupling poles. However, the nontrivial question
which we will now address is whether these poles can furthermore
become branch points and whether extra branch points can appear.
We are able to give a definite negative answer in some cases, and
expect this result to hold in general.

Branch cuts in the $\w$ plane may appear in $A(\w)$ and $B(\w)$ as
defined in (\ref{eq:seepole}). The analytic properties of these functions
are inhereted from the analytic properties of $\phi_\w(u)$ for small
$u$ as a function of $\w$. Recall that $\w$ specified the boundary
conditions as $u \to 1$. We shall shortly adapt some arguments from
\cite{newton} to prove some analyticity of $\phi_\w(u)$ in $\w$
\be\label{eq:strip}
-2 \pi < \Im\, \w \quad \Rightarrow \quad \text{analytic} \,.
\ee
Thus we gain a strip in the lower half plane in which $\phi_\w(u)$
is analytic and hence $G_R(\w)$ does not have branch points. However,
it was found in \cite{Starinets:2002br} that as $p$ becomes large, the
imaginary part of the location of the lowest quasinormal pole moves
towards the real axis. In particular, there exist infinitely many values
of $p$ for which the lowest quasinormal pole is within the
range (\ref{eq:strip}). The location of these poles will only be
altered by amounts of order $\a'$ and so will remain in
this strip. However, there can be no branch points
in this region and therefore these poles cannot have become branch points 
and
no new branch points can have appeared in this region. This is the
desired result. There does not seem to be a reason why these larger
$p$ poles should be special at strong coupling, so we expect that the
absence of branch cuts holds in general.

We first prove analyticity of $\phi_\w(u)$ in the region (\ref{eq:strip}) 
for the uncorrected potential. It is convenient to work in Regge-Wheeler
coordinates, so that equation (\ref{eq:bulkeqn}) assumes a
Schr\"odinger form. Set
\be
dr_* = \frac{1}{2\pi} \frac{du}{u^{1/2}(1-u^2)} \,,
\ee
which implies the range $0 \leq r_* < \infty$. Now let
\be
\psi = u^{-3/4} \phi \,.
\ee
The bulk equation (\ref{eq:bulkeqn}) becomes
\be\label{eq:schrod}
- \frac{d^2 \psi}{d r_*^2} + V(r_*(u)) \psi = \w^2 \psi \,,
\ee
where
\be
V(r_*(u)) = p^2 (1-u^2) + \frac{\pi^2}{u} \left[4 - \frac{(3 u^2 + 1)^2}{4} \right] \,.
\ee
The only property we need of this potential is the exponential decay towards 
the horizon
\be\label{eq:exponential}
V(r_*) \sim e^{-4\pi r_*} \quad \text{as} \quad r_* \to \infty \,.
\ee
The incoming boundary condition at the horizon (\ref{eq:incoming}) becomes
\be\label{eq:initial}
\psi_\w \sim e^{i \w r_*} \quad \text{as} \quad r_* \to \infty \,.
\ee

Now rewrite the Schr\"odinger equation (\ref{eq:schrod}) as an
integral equation incorporating the initial condition (\ref{eq:initial})
\be
\psi_\w(r_*) =  e^{i \w r_*} -
\frac{1}{\w} \int_{r_*}^{\infty} dx \sin \w (r_*-x) V(x) \psi_\w(x) \,.
\ee
The integral equation may be solved using a Born series
\be
\psi_\w = \sum_{n=0}^{\infty} \psi_\w^{(n)} \,,
\ee
where
\bea\label{eq:born}
\psi_\w^{(0)}(r_*) & = & e^{i \w r_*} \,, \nonumber \\
\psi_\w^{(n)}(r_*) & = & - \frac{1}{\w} \int_{r_*}^{\infty} dx
\sin \w (r_*-x) V(x) \psi_\w^{(n-1)}(x) \,.
\eea
Using the inequality
\be
\left| \sin\w(r_*-x) \right| \leq e^{- |\Im\, \w| (r_*-x)} \quad \text{for} 
\quad x > r_* \,,
\ee
it is straightforwardly shown that
\be
\sum_{n=0}^{\infty} \left| \psi_\w^{(n)}(r_*) \right| \leq
e^{-r_* |\Im\, \w|} \exp \left[\frac{1}{|\w|} \int_{r_*}^{\infty} dx
  |V(x)| e^{x (|\Im\, \w|-\Im\, \w)} \right] \,.
\ee
Therefore the Born series is absolutely convergent if
\be\label{eq:integral}
\int_{r_*}^{\infty} dx |V(x)| e^{x (|\Im\,\w|-\Im\,\w)} < \infty \,.
\ee
Given the exponential decay of the potential (\ref{eq:exponential}), it 
follows
that the integral (\ref{eq:integral}) is finite in the region 
(\ref{eq:strip}).
The consequent absolute convergence of the Born series
establishes continuity of $\psi_\w$ in this region. Analyticity follows
from performing an entirely analogous computation for the derivative of
$\psi_\w$ with respect to $\w$. Thus we prove the analyticity claim above 
for
the uncorrected case. Whilst the lower limit of integration, $r_*$, is
small, we may keep it finite and therefore do not need to worry about
divergences as $r_* \to 0$.

To extend this result to include $\a'$ corrections we need to perturb the 
solution and
the equation. In terms of the solution we have
\be
\psi \to \psi + \epsilon \psi' \,,
\ee
where $\epsilon$ is a small parameter controlling the $\a'$ corrections. The 
zeroth order solution
$\psi$ satisfies (\ref{eq:schrod}) whilst the perturbation satisfies
\be
- \frac{d^2 \psi'}{d r_*^2} + V(r_*(u)) \psi' + {\mathcal{O}} \psi = \w^2 
\psi' \,,
\ee
where ${\mathcal{O}}$ is a higher order differential operator which contains 
the effects of the
$\a'$ corrections. This equation is the same as that satisfied by the zeroth 
order
solution but with an extra inhomogeneous term. We do not need to know the 
precise
form of ${\mathcal{O}}$,
but note that consistency of the $\a'$ expansion requires
\be\label{eq:consistency}
|V \psi'|, |{\mathcal{O}} \psi| \ll \frac{1}{\epsilon}
|V \psi| \qquad \text{as} \quad r_* \to \infty \,.
\ee
The integral equation satisfied by the perturbation is
\be
\psi'_\w(r_*) =  e^{i \w r_*} -
\frac{1}{\w} \int_{r_*}^{\infty} dx \sin \w (r_*-x)
\left[V(x) \psi'_\w(x) + {\mathcal{O}} \psi_\w(x) \right] \,.
\ee
We may again solve this equation using a Born series. All the terms are
as previously in (\ref{eq:born}) except for the first term which is now
\be
\psi'{}_{\w}^{(0)}(r_*) = e^{i \w r_*} - \frac{1}{\w} \int_{r_*}^{\infty} dx 
\sin \w (r_*-x)
{\mathcal{O}} \psi_\w(x) \,.
\ee
The zeroth order solution, $\psi_\w(x)$, satisfies the inequality
\be\label{eq:bound}
\left| \psi_\w(x) \right| \leq k e^{-x \Im\,\w} \,,
\ee
where $k$ is some constant.
The bound (\ref{eq:bound}) follows from the asymptotic behaviour of
$\psi_\w(x)$ (\ref{eq:initial}) and the absence of divergences at finite 
values of $x$ and as $x \to 0$. Using this bound together with the inequalities 
(\ref{eq:consistency})
it follows that
\be
\sum_{n=0}^{\infty} \left| \psi'{}_{\w}^{(n)}(r_*) \right| \leq
e^{-r_* |\Im\,\w|}
\left(
\left[ 1+ {\textstyle \frac{k}{\epsilon}} \right]
\exp \left[\frac{1}{|\w|} \int_{r_*}^{\infty} dx |V(x)| e^{x 
(|\Im\,\w|-\Im\,\w)}\right]
- {\textstyle \frac{k}{\epsilon}} \right)
\,.
\ee
Thus if (\ref{eq:integral}) holds, the Born series is absolutely convergent. 
Along
with an analogous result for the derivative of $\psi'_{\w}$ with respect to 
$\w$,
the convergence again implies analyticity of $\psi'_{\w}$ in the strip 
(\ref{eq:strip}).
This completes the proof of our claim including $\a'$ corrections.

\section{Discussion}

We have seen that frequency space glueball propagators in large $N$
thermal ${\mathcal{N}} = 4$ SYM theory on $\R^{1,3}$ have different analytic
structures at weak and strong 't Hooft coupling. In position space this
difference translated into qualitatively different time decays of
field fluctuations about equilibrium. In the perturbative regime the
correlators are known to exhibit power law falloff on intermediate
time scales followed by a slow exponential decay at late times. We
contrasted this with the rapid pure exponential decays seen at strong
coupling. 

Based on our free field computations and the generally understood
aspects of weakly interacting plasmas, we expect that the
perturbative real time response functions will exhibit branch cut
non-analyticities. These differ qualitatively from the strong coupling
quasinormal poles, which we further argued survive away from the
strict infinite coupling limit. This picture suggests that these
propagators must depend non-analytically on the 't Hooft
coupling. The non-analyticity could either be due to nonperturbative
corrections to both weak and strong coupling limits or else to a phase
transition at an
intermediate coupling. As we noted, however, further work is needed to
rule out the possibility that the branch cuts seen in perturbation
theory actually split into a series of poles in the large $N$ theory,
thus allowing a continuous interpolation to strong coupling.

One observation that may favour the phase transition possiblity is
that to each order in the large $N$ expansion, the perturbative 't
Hooft expansion for zero temperature ${\mathcal N} = 4$ SYM theory
is thought to be regular at $\lambda = 0$, with a finite radius of
convergence. This is due to the slow growth of the number of planar
diagrams at each order in $\lambda$ \cite{Brezin:1977sv}. Including a finite
temperature should not modify this regularity beyond introducing a
dependence on $\sqrt{\lambda}$. The convergence of the 't Hooft expansion does not rule
out nonperturbative terms in $\lambda$, but it does mean that they
are not necessary, unlike in standard Yang-Mills theory. If the
non-analyticity is not due to nonperturbative corrections, then our
results predict a phase transition in finite temperature Yang-Mills
theory on $\R^{1,3}$.

Regarding corrections to the strong coupling results, we have argued
that perturbative $\a'$ corrections do not lead to a broadening or
merging of the poles into branch cuts. In principle there could be
nonperturbative $\a'$ corrections to the strong coupling results, but
there are no obvious worldsheet instantons in the bulk. Our strong
coupling considerations have all been in the strict large $N$ limit.
It is possible that the quasinormal poles are broadened into
branch cuts by $1/N$ effects and it
is expected that $1/N$ corrections will introduce additional branch
cuts in real time correlators at all couplings
\cite{Kovtun:2003vj}.

A natural candidate for a dual phase transition in the bulk is given by
the Horowitz-Polchinski correspondence
\cite{Horowitz:1996nw, Susskind:1993ws}. As the AdS background becomes
increasingly curved, the horizon size eventually becomes of the order
of the string scale. The Horowitz-Polchinski correspondence suggests
that in this regime the appropriate description of the system is as an
ensemble of highly excited string states. It would be interesting to
ascertain whether there is a connection between the qualitative change
in behaviour we find in field theory from strong to weak coupling and
the putative change in the bulk from a black hole to string states.

These possibilities have been entertained before.
A phase transition at a critical 't Hooft coupling in finite
temperature ${\mathcal{N}} = 4$ SYM theory was previously argued for in
\cite{Li:1998kd}, mainly on the basis of analytic properties of the
free energy as a function of coupling. It was then suggested that this
phase transition was dual to the Horowitz-Polchinski point in
\cite{Gao:1998ww}\footnote{This paper further argued
that the Hawking-Page transition would not continue to weak
coupling; but this now appears not to be the case
\cite{Sundborg:1999ue,Aharony:2003sx}. However,
that argument seems to be independent of the
arguments in the same paper and in \cite{Li:1998kd} for a phase
transition as a function of the 't Hooft coupling.}.
A key question for placing the arguments about phase transitions on a
firmer footing is the identification of an appropriate order
parameter. One would also want to establish whether the phase
transition existed only in dynamical response functions or whether the
static equilibrium theory was also affected.

Finally, in a different direction,
various proposals have been made for isolating features of the bulk
black hole singularity in dual field theory correlators
\cite{Fidkowski:2003nf,Festuccia:2005pi}. These works have focused on
infinitely massive bulk probes in order to employ a semiclassical
approximation. The correlators we have computed in field theory are dual to 
massless minimally coupled bulk scalars. Nonetheless, it would be
interesting to 
see if signatures of a singularity, possibly resolved, can be seen in
our expressions.

\vspace{0.5cm}

{\bf Acknowledgements}: During this work we have had useful discussions
and correspondence with Nick Dorey,
Michael Green, Shiraz Minwalla, 
Asad Naqvi, Toby Wiseman and Larry Yaffe. We are
especially grateful to Ofer Aharony, Dan Boyanovsky and Rob Pisarski
for critical comments and extensive discussions.
SAH is supported by a research fellowship
from Clare College, Cambridge. SPK acknowledges support from a
PPARC Rolling Grant.

\appendix

\section{Effect of thermal mass and width}

We discuss how the inclusion of the thermal mass
$m_T\sim\sqrt\lambda T$ and thermal width $\Gamma$
for the intermediate gluon
states in the correlator (\ref{eq:ptfn}) changes the analytic
structure of the frequency space propagator. Including a mass
changes the frequencies in the real time propagators (\ref{eq:freept}) to
$E=\w_p \equiv \sqrt{p^2+m_T^2}$. Restricting attention to the homogeneous
case (the inhomogeneous case is similar) the retarded glueball
propagator becomes
\be
\langle \En^0(t) \En^0(0) \rangle_R = -4 {N^2\over \pi^2} \Theta(t)
\int_0^\infty p'^2 dp'{(2p'^2+m_T^2)^2\over \w_{p'}^2}
\coth\left({\w_{p'}\over 2 T}\right) \sin(2 \w_{p'}t)\,,
\ee
where $\w_{p'}=\sqrt{p'^2+m_T^2}$. Upon Fourier transforming to
frequency space this function is seen to have branch
cuts. The easiest way to see this is to note that upon Fourier
transforming  $\Theta(t)\sin(2\w_{p'}t)$ using an $i\epsilon$
prescription we get 
\be
\label{eq:massive}
G_R^0(\w)= 2 {N^2\over \pi^2}\int_0^\infty p'^2
dp'{(2p'^2+m_T^2)^2\over \w_{p'}^2} 
\coth\left({\w_{p'}\over 2 T}\right)
 \left[{1\over{\omega-2 \omega_{p'}+i\epsilon}}
   -{1\over{\omega+2 \omega_{p'}+i\epsilon}}\right]\,.   
\ee
The presence of discontinuities can now be inferred from the imaginary
parts of the energy denominators in the above expression by using the
representation
\be
{1\over{\omega\pm 2 \omega_{p'}+i\epsilon}}=
P{1\over{\omega \pm 2 \omega_{p'}}}+i\pi\delta(\omega \pm 2 \omega_{p'})\,. 
\ee
It is now a straightforward excercise to see that (\ref{eq:massive})
has a square root branch cut discontinuity between $\w=2 m_T$ and
$\w=-2m_T$. 
%It is also possible to show that (\ref{eq:massive}) has
%branch 
%cuts as we move down the negative imaginary 
%axis between $\w = -2 m_T - 4 \pi i T n$ and $\w=2m_T - 4\pi i T n$.
Note that this cut on the real axis would also appear for a massive
particle at zero temperature.

Finally, we turn to the inclusion of the thermal width in the
propagators. This is a subtle issue since it involves unstable
intermediate states. We can provide a heuristic argument for its
effect by simply looking at the form of the propagators for a massive
but unstable resonance at zero temperature. The propagator for such a
resonance, at zero temperature, is $(p_0^2-p^2 -m^2 +i\Gamma)^{-1}$
where $\Gamma$ is the width of the resonance. Repeating the above
excercise in the real time domain and computing the frequency space
retarded propagator, it is clear that the branch cut along the real axis
gets shifted off the real axis. Whether this conclusion is correct 
depends on a more detailed and precise field theoretical analysis  of
the analytic structure of self-energy corrections to the internal lines.

\section{Thermal glueball propagators on $\R \times S^3$}

In this appendix we indicate how our field theory computations may be 
generalised to
thermal Yang-Mills theory on $\R \times S^3$. The glueball time
dependence in this context is interesting as it may be a formalism in
which one can study nonequilibrium processes such as black hole
evaporation.

\subsection{Expansion in harmonics}

Free Yang-Mills theory on $\R \times S^3$ has a nontrivial phase
structure because the finite spatial volume restricts the partition
function to a sum over colour singlets \cite{Sundborg:1999ue,
  Aharony:2003sx}. As emphasised in \cite{Aharony:2003sx} this may be
seen in the fact that the nondynamical mode
\be
\a^a \equiv \frac{g}{\text{Vol}S^3} \int_{S^3} A^a_0\, d\Omega \,,
\ee
is strongly coupled even in the $g^2 N \to 0$ limit and so needs to be 
treated
nonperturbatively. The different phases are characterised by the
distribution of eigenvalues of $\a$ \cite{Aharony:2003sx}.

Working on $S^3 \times \R$, it is useful to decompose the
gauge potential into transverse vector spherical harmonics on $S^3$
\be\label{eq:decomp}
A_i^a(t,\q) = \sum_m A^{a m}(t) V_i^m(\q) \,, \nonumber \\
\ee
In this notation, $m$ denotes all the relevant quantum numbers for the
harmonic decomposition. For scalar harmonics these are $\{ j_m, m_m, n_m\}$
and there is an extra $\e^m = \pm 1$ for the vector harmonics
\cite{Aharony:2005bq, Cutkosky:1983jd}. The quadratic part of the
Yang-Mills action then becomes
\be\label{eq:harmonicaction}
\Lag_2 = -\frac{1}{2} A^{a \bar{m}} \left[D_t^2 + E_m^2 \right] A^{a m} \,.
\ee
As previously, in free field we may work directly with the physical
transverse degrees of freedom and ignore the other modes and ghosts.
In expression (\ref{eq:harmonicaction})
\be
D_t A^{a m} = \frac{\pa A^{a m}}{\pa t} + f^{abc} \a^b A^{c m} \,,
\ee
and the energy levels are
\be
E_m = j_m+1 \,.
\ee
We are working in units in which the radius of the $S^3$ is one.

The glueball operator (\ref{eq:simpleglueball}) may also be expanded
into harmonics. We restrict ourselves here to the $\Tr F^2$
operator. The quadratic term gives
\be\label{eq:glueball}
\En_2 = 2 \left[- D_t A^{am} D_t A^{an} D^{mnp} +
  \e^m \e^n E_m E_n A^{am} A^{an} D^{mnp} \right] S^p \,.
\ee
This expression should also contain terms dependent on the time
component $A_0$ of the gauge field, but these do not contribute to the
free theory propagators.
The expression (\ref{eq:glueball}) is obtained using completeness of
the harmonic decomposition and the triple overlap \cite{Cutkosky:1983jd}
\be\label{eq:triple}
D^{mnp} = \int_{S^3} d\Omega \, V^{i m} V^n_i S^p \,,
\ee
where $S$ and $V$ are the scalar and vector harmonics, respectively.

The thermal propagator following from the action
(\ref{eq:harmonicaction}) is now, for $t > 0$,
\be\label{eq:freeprop}
\langle A^{am}(0) A^{bn}(t) \rangle  = \d^{mn}\frac{i e^{-f \a t}}{2 E_m}
\left[(\bar{n}(E_m)+1) e^{-i E_m t} + n(E_m)
e^{i E_m t} \right] \,,
\ee
where we have suppressed the colour indices and used the
abbreviation
\be\label{eq:falpha}
\left(f \a\right)^{a c} \equiv f^{a b c} \a^b \,.
\ee
The Bose-Einstein distribution is now
\be\label{eq:be}
n(E) = \frac{1}{e^{\b(E-if\a)}-1} \,.
\ee
We can see in (\ref{eq:be}) that the $\a$ condensate plays the role
of a chemical potential. This
corresponds with the fact that $\a$ is a Lagrange multiplier for the
conservation of colour charge.

\subsection{The free theory two point function}

We find that for $t > 0$ the two point function of the glueball 
(\ref{eq:glueball})
is given by
\be\label{eq:zeroorder}
\langle \En^p(0) \En^q(t) \rangle = - \frac{\d^{p,\bar q}}{2\pi^2 E_p}
\Tr \, \left[\sum_{I_+} c^+_{mnp} \,
  Q^+_{mn}(t)  + \sum_{I_-} c^-_{mnp}
  \, Q^-_{mn}(t) \right] \,,
\ee
where the first sum runs over
\be
I_+ = \{ m,n \mid |j_m-j_n| \leq j_p \leq j_m + j_n -2 \} \,,
\ee
and the second over
\be\label{eq:I}
I_- = \{ m,n \mid |j_m-j_n| + 2 \leq j_p \leq j_m + j_n\} \,.
\ee
The coefficients are
\bea
c^+_{mnp} & = &  \left[(E_m + E_n + E_p)^2-1\right]
\left[(E_m + E_n - E_p)^2-1\right] \,, \nonumber \\
c^-_{mnp} & = & \left[(E_m + E_p -
    E_n)^2-1 \right] \left[(E_n+E_p-E_m)^2-1 \right] \,,
\eea
and the time dependence is contained in
\bea\label{eq:Q}
Q^+_{mn}(t) & = &  (\bar{n}(E_m)+1)(n(E_n)+1) e^{-i(E_m + E_n)t}
+ n(E_m) \bar{n}(E_n) e^{i(E_m + E_n)t} \nonumber \,, \\
Q^-_{mn}(t) & = & (\bar{n}(E_m)+1)\bar{n}(E_n)e^{-i(E_m - E_n)t}
+ n(E_m) (n(E_n)+1) e^{i(E_m - E_n)t} \,.
\eea
In deriving these results one needs to use identities for 3j-symbols
that may be found for example in the appendices of
\cite{Aharony:2005bq}.

The trace in (\ref{eq:zeroorder}) is over the colour indices of
$f\a$ as given in (\ref{eq:falpha}). This may be expressed in
terms of the eigenvalues $\lambda_i$ of $\a$ considered in
\cite{Aharony:2003sx} as follows: For some function $F$ of $f
\a$ we have
\be
\Tr F(f \a) = \sum_{i,j=1}^N F(\l_i - \l_j) \,.
\ee
In practice, we need to evaluate this sum in the two possible vacua 
identified
by \cite{Aharony:2003sx}. The first is the maximally clumped vacuum in which
all of the $N$ eigenvalues are equal
\be
\l_i = \text{const.} \quad \Rightarrow \quad \Tr F(f \a) = N^2
F(0) \,.
\ee
This is the correct vacuum at high temperatures.
The other possibility is a uniformly distributed vacuum
\be\label{eq:uniformtrace}
\l_i = \frac{2\pi i}{N} \quad \Rightarrow \quad \Tr F(f \a) = 2N
\sum_{n=1}^{N-1} F\left(\frac{2\pi n}{N}\right) \left(1-\frac{n}{N}\right) + 
N F(0) \,,
\ee
which is the correct vacuum at low temperatures.
In this last expression we assumed that $F$ is an even function
which is indeed true for (\ref{eq:zeroorder}).

The two point function (\ref{eq:zeroorder}) has no mixing
between the different $S^3$ harmonics, as we expect for a free
theory. The simplest case to consider is the propagation of the
homogeneous mode $p=q=0$. In this case from (\ref{eq:I}) we have
$E_m=E_n$ in (\ref{eq:zeroorder}) and only the first sum contributes
to the time dependence. For the $\l_i = \text{const}$ vacuum
the sums in (\ref{eq:zeroorder}) for the homogeneous mode
admit a compact expression in terms of theta functions
\bea\label{eq:theta}
\Re \langle \En^0(0) \En^0(t) \rangle & = & \frac{-N^2}{2\pi^2} \left[-4
  \frac{d}{dt} + \frac{d^3}{dt^3} \right] \frac{d}{d\b}
\frac{\vartheta'_1(t,e^{-\b/2})}{\vartheta_1(t,e^{-\b/2})} \,,
\nonumber \\
\Im \langle \En^0(0) \En^0(t) \rangle & = & \frac{-N^2}{2\pi^2}
\left[ 4 \frac{d^2}{dt^2} - \frac{d^4}{dt^4} \right] \left[
\frac{\vartheta'_1(t,e^{-\b/2})}{\vartheta_1(t,e^{-\b/2})} -
\frac{\vartheta'_1(t,e^{\b/2})}{\vartheta_1(t,e^{\b/2})} \right] \,.
\eea
These expressions are true for $t>0$. We have dropped various $\d(t)$
contributions to the real part. We have not attempted to generalise
(\ref{eq:theta}) to the inhomogeneous modes.

One significant feature of the result given in expressions
(\ref{eq:zeroorder}) to (\ref{eq:Q}) is that it is periodic in time.
The periodicity of (\ref{eq:zeroorder}) under $t \to t + 2\pi$
is a special case of the quasi-periodicity that one
always finds for field theories on a compact spatial manifold. The exact
periodicity in this case arises because in our free theory on $S^3 \times 
\R$
the energy eigenvalues are evenly spaced. The quasi-periodicity is
expected to be dual to large gravitational fluctuations in the bulk
\cite{Maldacena:2001kr} such as the formation and evporation of black
holes. However, this correspondence has only been made precise when
averaged over long times \cite{Barbon:2003aq,Barbon:2004ce}. The
requirement of long time averaging is due to the thermal nature of the
semiclassical approximation in gravity.

At increasingly high temperatures, $\b \to 0$, the nontrivial time
dependence is restricted to times $t \lesssim \b$, much shorter than
the full period. This limit, which can be viewed as letting $S^3
\to \R^3$, is simpler to study analytically was the subject of the
prvious three sections. Note however that the nontrivial phase structure 
uncovered in
\cite{Sundborg:1999ue,Aharony:2003sx} occurs at finite temperatures. In
particular, addressing black hole evaporation in this formalism will
require working at finite temperature.

At low temperatures, $\b \to \infty$, we need to use the vacuum with
evenly spaced eigenvalues \cite{Aharony:2003sx}. Thus we use the formula
(\ref{eq:uniformtrace}) to evaluate the traces. In particular,
evaluating the leading finite temperature correction will require
the result
\be
\Tr \cos \beta f \a = \frac{\sin^2 \beta \pi}{\sin^2 \frac{\beta
    \pi}{N}} \,.
\ee
The resulting two-point correlator is then
\be
\langle \En(0) \En(t) \rangle = - \frac{i 2 N^2}{\pi^2 \sin^5 t}
\left[\cos t+2\cos^3 t \right] - \frac{\sin^2 \beta \pi}{\pi^2 \sin^2 
\frac{\beta
    \pi}{N}} e^{-2 \b} e^{-i 4 t} + \cdots \,.
\ee
The retarded propagator is then
\be
\langle \En(0) \En(t) \rangle_R = - \frac{4 N^2}{\pi^2 \sin^5 t}
\left[\cos t+2\cos^3 t \right] + 2 e^{-2\b} \frac{\sin^2 \beta \pi}{\pi^2 
\sin^2 \frac{\beta
    \pi}{N}} \sin 4 t + \cdots \,.
\ee
The leading finite temperature correction is exponentially suppressed.

\end{document}